# Power Area Density in Inverse Spectra

*Matthias Rang[1] and Johannes Grebe-Ellis[2]*
[1]*Forschungsinstitut am Goetheanum, Dornach*
[2]*Bergische Universität Wuppertal*

**Abstract**

In recent years, inverse spectra were investigated with imaging optics and a quantitative description with radiometric units was suggested (Rang 2015). It could be shown that inverse spectra complement each other additively to a constant intensity level. Since optical intensity in radiometric units is a power area density, it can be expected that energy densities of inverse spectra also fulfill an inversion equation and complement each other. In this contribution we report findings on a measurement of the power area density of inverse spectra for the near ultraviolet, visible and the infrared spectral range. They show the existence of corresponding spectral regions *ultra-yellow* (UY) and *infra-cyan* (IC) in the inverted spectrum and thereby present additional experimental evidence for equivalence of inverse spectra beyond the visible range.

**Introduction**

In his theory of colour, Goethe described the observation that inverse optical contrasts, viewed through a prism, lead to complementary spectral phenomena (Goethe 1970, Müller 2015).[1] This observation has been variously addressed and linked to the questions of whether the inverted spectrum can be physically applied (Kirschmann 1917, 1924) and whether the complementarity of spectral phenomena can be generalized experimentally (Bjerke 1961). Particular attention has been given to the problem of inverting Newton's *Experimentum crucis*, designed to demonstrate the purity of selected spectral regions (Holtsmark 1970, Sällström 2010). The philosophical discussion on Goethes theory of colour has been rekindled since Olaf Müller published his comprehensive work on Goethe, Newton and the controversy about colour theory (Müller 2015). He claimed that Goethe's discovery of the complementarity of spectral phenomena had been overlooked and that in particular his symmetry argument had not been experimentally tested by physicists. A study of the physical conditions under which the complementarity of spectral phenomena can be generalized was carried out recently (Rang 2015). It was shown that complementarity as a symmetry property of inverse optical spectral phenomena is given when the optical system can be viewed as a system with conserved radiation energy (Rang et al. 2017).[2]

---

[1] Optical contrasts, which merge into one another by reversing their luminance, are called *inverse*. Then $L(x,\vartheta,\varphi) + \bar{L}(x,\vartheta,\varphi) = $ const holds true for the luminance $L$ and the reversed luminance $\bar{L}$ for all positions $x$ with the same constant (Rang 2015). The inversion of spectra can be equivalently defined. To specify this definition for irradiance instead of luminance has the advantage of being consistent with the units of the measured values. Inverse spectra add up to a wavelength-independent, but not necessarily white irradiance standard. In the case of a broadband illumination, which includes at least the entire VIS region, the irradiance standard becomes white and one speaks of *complementary* spectra. Therefore, complementary spectra are a subclass of inverse spectra.

[2] The conservation of radiation energy plays a key role also in the proof of Babinet's theorem. According to it the inversion of diffraction structures (as for example slit apertures) leads to *identical* diffraction images. The theorem applies only to diffractive structures as *imaging* elements. In contrast, slit apertures are not used as *imaging* elements but as *imaged* elements in spectroscopical applications. In the latter case inverse structures result in *inverse* images, irrespective of whether the aperture is imaged by a prism, a grating or an inverted grating.

In the past, inverse spectra have been described and discussed only as phenomena in the visible domain (VIS). As to the question of whether the symmetry properties of inverse optical spectra extend beyond the visible part of the spectrum into the infra-red (IR) and ultra-violet (UV) range we have only been able to find a short note by Kirschmann (Kirschmann 1917). He reported a photographic blackening in the region of the inverted spectrum that corresponds to the UV region in the ordinary spectrum and he pointed out that a region corresponding to the IR could also be expected. The preliminary work on the complementarity of the spectra and their geometric isomorphism give rise to the well-founded assumption that a corresponding measurement of complementary spectra yields inverse power densities for arbitrarily selected spectral regions. The results of such a measurement are presented in the present paper. They confirm the abovementioned assumption, and in particular show: 1. The power density of the inverse spectrum is less than that of its bright background, which means that it becomes negative when the measurement is referred to the intensity level of this background; 2. The infrared region (IR), which adjoins the long-wavelength region of the visible spectrum corresponds to an invisible *infra-cyan* region (IC) in the inverted spectrum; 3. The ultra-violet region (UV) adjoining the short-wavelength region of the visible spectrum corresponds to an invisible *ultra-yellow* region (UY) in the inverse spectrum. Following the presentation of the experiment and the discussion of the results, we give a brief discussion about the relevance of energy conservation for the generation of inverse spectra.

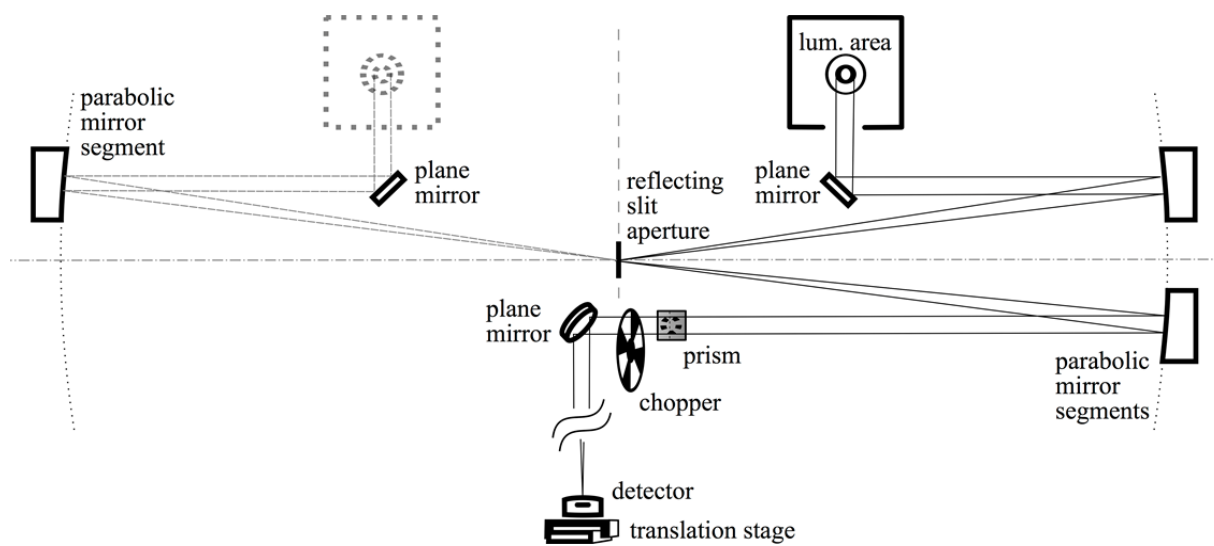

**Fig. 1:** Experimental setup. The plane of symmetry in the setup (dashed line) is defined by the reflecting slit aperture. The rotational axis of the off-axis parabolic mirror segments (dot-and-dash line) lies perpendicular to this plane. From the optical design (concatenated beam path) it follows that the real image of the luminous area of the light source is located centered in the prism (dotted circle). By "mirroring" the light source at the plane of symmetry, the change from the inverse (solid lamp housing) to the ordinary spectrum (dotted) is carried out. The horizontal orientation of the slit and the vertical refraction plane of the prism cause the plane of beam paths to be tilted after passage through the prism.

**Experimental Design**

The measurements were carried out with a spectroscopic setup which was especially developed for the generation of inverse spectra with wide spectral ranges. The components of the setup are grouped symmetrically around a reflecting slit aperture at which contrasts occur in transmission *and reflection* that are mutually inverse (Fig. 1). The function of the aperture, which results in both transmission and reflection, is similar to that of a beamsplitter or interference filter, and thus permits the simultaneous generation of inverse spectra under approximately identical experimental conditions (Rang & Grebe-Ellis 2009). We illuminated this aperture via an off-axis parabolic mirror element with a xenon high pressure lamp (LOT Oriel LSB 1000W).

The reflecting aperture is imaged by a further off-axis parabolic mirror segment through a prism (non-coated 30° N-BK7-Littrow-prism) at a distance of 7.5m on a projection screen. The design is based on a concatenated beam path, which creates a real image of the luminous area inside the prism. As a result, the dispersive effect of the prism on the image can be eliminated.[3]

Apart from the prism, all components were designed as reflection optics with surface aluminum coatings in order to allow a spectral range from approximately 300nm up to 2.8μm in the experiment. In the spectrum a pyroelectric radiation energy sensor (Ophir RM9) was positioned on a translation stage with 750mm of travel length (Igus SHT-12 linear module). The sensor was narrowed to a width of 3mm by means of a slit aperture and was thus matched to the aperture width of the reflecting slit (300μm), which was enlarged about ten times in the spectrum. To improve the signal-to-noise ratio, a chopper was inserted in the beam path and only the detector signal modulated with the chopper frequency (18Hz) was selectively amplified (lock-in amplifier). The spectroscope's bandpass, which is composed of the slit width, the wavelength-dependent image magnification of the slit image and the dispersion-dependent resolution limit of the prism, is theoretically 2.6nm at the detector position 0mm (corresponding to 316nm) and 114nm at the position 300mm (corresponding to 2828nm).

**Results**

The spectra were recorded by the detector as a radiation flux. 300 individual measurement values were recorded and averaged to a mean value at each of 150 detector positions. The typical standard deviation of the mean values was statistically calculated to approximately 0.84nW (0,04nW/mm$^2$). However, the accuracy of the detector is only 3% of the reading. In this case the error is caused mainly by the detector´s wavelength-dependent inaccuracy and especially by the drift of the emission power of the light source. The spectral background resulting from detector-noise and scattered light in the experiment was similarly measured at the borders of the measuring range on both sides and subtracted from the measured values over the entire measuring range. The measured values thus freed from the background were multiplied by the spectral sensitivity of the detector and converted into irradiance by normalizing it with the area of the detector surface.

The irradiance in the ordinary spectrum is shown in Figure 2 as a function of the position of the detector on the imaging screen. The corresponding wavelengths are indicated in the upper horizontal axis. The nonlinearity of the wavelength axis results from the nonlinear dispersion of the prism material.[4] The diagram shows the spectral distribution of the xenon high pressure lamp. In addition to the characteristic peaks at wavelengths of around 470, 840, 910, 1000 and 1500nm, it has a continuous spectrum similar to the spectrum of a black body.[5] A photograph of the spectrum is shown below the spectral profile. The visible range of this spectrum registered by the camera extends from about 430nm to 690nm and corresponds well to the visible spectrum.[6]

---

[3] If the mirror slit aperture (Fig. 1) – and thus the effective contrast element – is replaced by a perfect mirror, the projection screen is at any point homogeneously illuminated and shows no wavelength dependency over the full measuring range. Besides the mirror slit aperture the mountings of the optical elements are also contrasts. However, these do not result in a spectral decomposition on the screen, if the positioning is correct in the concatenated beam path. Thus through the prism, the screen is illuminated just as it is with removed prism, except for the reflection losses at the glass surfaces (Rang 2015).

[4] The wavelength allocation between detector position and spectral component was calculated by means of the Sellmeier-equations with coefficients of BK7 and the deviation angles at the prism and then calibrated by a known xenon peak at approximately 840nm (Osram datasheet) and tested by measured positions of five known mercury emission lines.

[5] The ultra-violet range of the spectrum is suppressed due to the lower reflection of the aluminum coatings in the ultra-violet. This is particularly noticeable here due to the fivefold reflection on the optical components (Fig. 1).

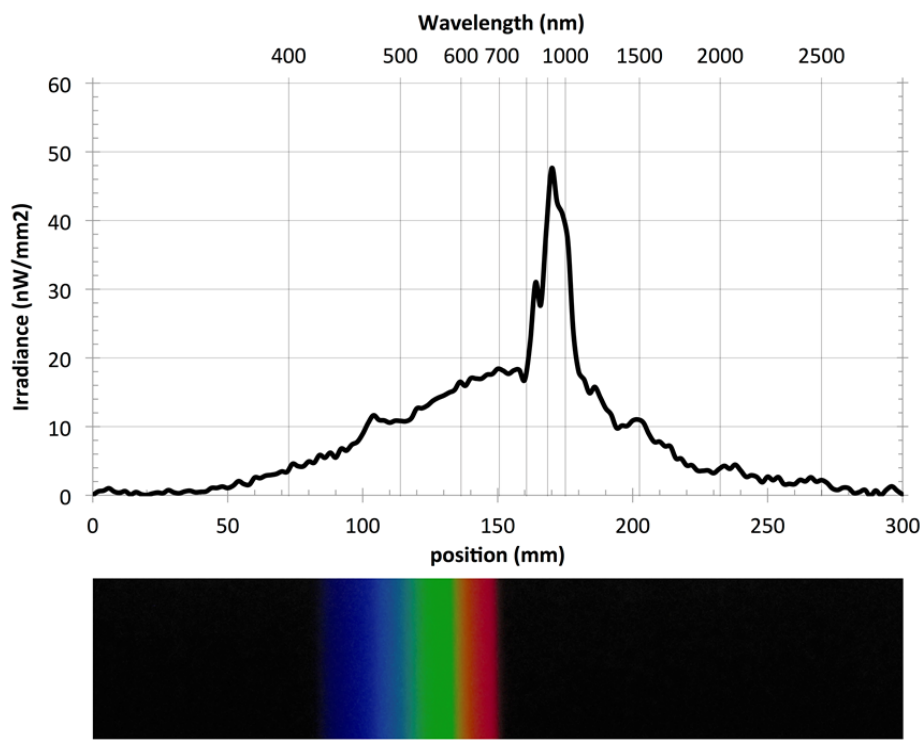

**Fig. 2:** Irradiance of the xenon high pressure lamp in the ordinary spectrum after subtraction of the background, including a photograph of the spectrum itself.

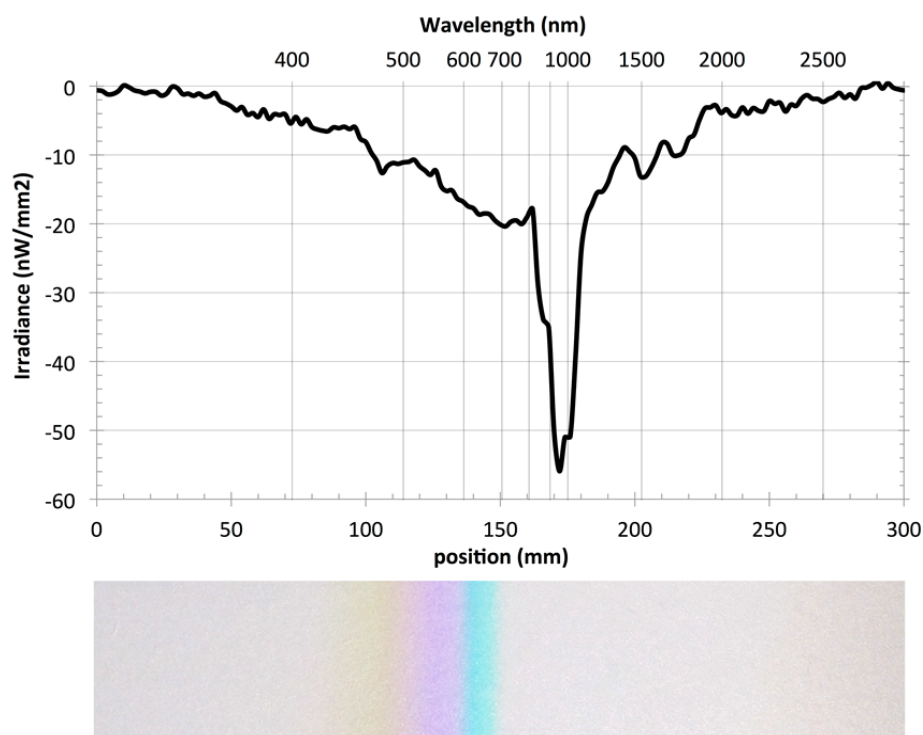

**Fig. 3:** Irradiance of the xenon high pressure lamp in the inverted spectrum after subtraction of the background, including a photograph of the spectrum itself.

To measure the corresponding inverse spectrum, the xenon lamp was moved on a rail to its mirrored position, so that the reflecting aperture was illuminated from the opposite side (Fig. 1). Thus the transmission and reflection beam path were interchanged in the experiment with the effect that the inverse spectrum is generated by the same optical components and congruent to the ordinary spectrum. This ensured that both spectra were measured under virtually the same experimental conditions.

---

[6] The extent of the visible spectrum can be estimated by means of the light sensitivity curve of the eye, which has its maximum for colour and daylight in the green spectral range at approximately 555nm and decreases in the red and violet region of the spectrum. At 430nm and at 690nm it is only about 1% of its maximum value.

The registered radiation flux was converted into irradiance as described above.[7] It is particularly noteworthy that the bright background of the spectrum were measured in the border areas of the measuring range and treated as an ordinary spectral background.[8] Figure 3 shows the resulting irradiance in the inverse spectrum as function of the detector position on the projection screen. The corresponding wavelengths are again indicated at the upper horizontal axis and a photograph of the spectrum has been added in the measuring range at the bottom.

By subtracting out the background of the inverse spectrum, all shown irradiances become negative. The ordinate axis with its zero point above thus represents a deviation of the irradiance from this background. In comparison, both spectra have similar but inverse spectral profiles.

**Discussion**

The results verify quantitatively the invisible spectral regions of the inverse spectrum for the first time. Adjoining the yellow border of the inverse spectrum at 430nm we found an *ultra-yellow* spectral region (UY) which extends to approximately 350nm in this case. Similarly, the cyan region of the visible spectrum is adjoined by an *infra-cyan* region (IC) with negative peaks and details as found in the corresponding IR region. The existence of these spectral regions demonstrates empirically that the inverse spectrum is not merely a visual phenomenon and should therefore not be considered as only a special feature of the visible part of the entire optical domain.

The optical equivalence of the two spectra requires, however, that they not only have corresponding spectral regions, but also fulfill a quantitative inversion equation over the entire spectral range. This is the case if the equation

$$E^{\mathrm{r}}(x) + \bar{E}^{\mathrm{r}}(x) = 0 \quad \forall x \qquad (1)$$

holds true for all spectral positions $x$, where $E^{\mathrm{r}}(x)$ is the referenced irradiance in the ordinary spectrum, and $\bar{E}^{\mathrm{r}}(x)$ is the referenced irradiance in the inverse spectrum. As "referenced irradiance" we denote an irradiance after it has been freed from its spectral background by subtraction (Rang 2015).[9] It is noticeable that the irradiance in the ordinary spectrum is on average between 10% and 15% lower than in the inverse spectrum. This is probably due to a geometrically narrowed gap-width in transmission caused by the finite thickness of the reflecting aperture body.

In addition to the measured irradiance of the two spectra (Fig. 4: light grey solid lines), their sum (Fig. 4: dark grey solid line) was calculated after the inverse spectrum has been scaled to 90% (dark grey dashed line) in order to take account of the geometrically constricted width of the slit in the ordinary spectrum.

---

[7] Due to the higher background and the increased irradiance of the inverse spectrum, a slightly higher typical standard deviation of the mean value of 1.04nW (0.04nW/mm$^2$) is calculated here.

[8] In both the ordinary and the inverse spectrum the background was not quite flat, so that the subtraction in both cases led to a small change of the spectral profiles. This is due to an inhomogeneous angular emission of the light source. The emission of the light source is affected at moderate opening angles by a decrease of the luminous area caused by perspective and minimal shadowing of the arc caused by the electrodes of the xenon lamp. Additionally the angular dependency of reflection loss at the prism and inhomogeneous properties of optical coatings may have an effect on the background. In the inverse spectrum, however, the illumination inhomogeneity was significantly higher in the measuring range.

[9] Mathematically, an element $b$ is called a (two-sided) inverse element to $a$ when $a \circ b = b \circ a = e$, where $e$ is the neutral element of the binary operation. In the present case of an additive operation $\bar{E}^r$ is designated as an additive inverse element to $E^r$ with the neutral element 0. From the commutativity of the additive operation it follows that $E^r$ is also the additive inverse element to $\bar{E}^r$. In addition to this mathematical analogy of inverse elements in sets and inverse spectra, the above inversion equation can also be derived with a radiometric formalism from optical considerations (Rang 2015).

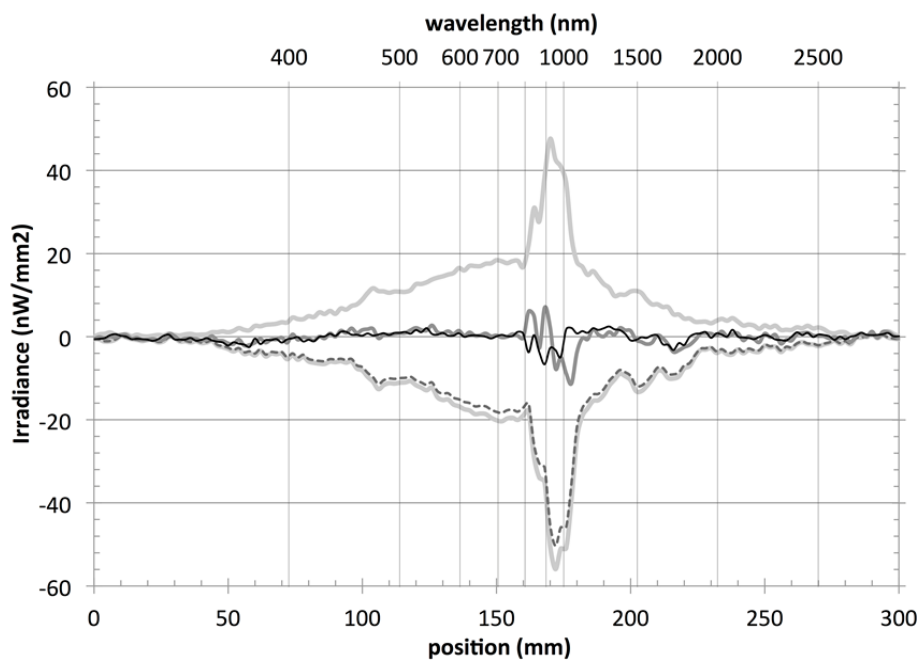

**Fig. 4:** Irradiances shown in Fig. 2 and 3 (light grey) and the corrected irradiance in the inverse spectrum (dark grey dashed line). The dark grey solid line displays the sum of the corrected irradiances. The fine black line shows the sum in the case of an additional correction of the displacement between both spectra.

Figure 4 shows that the illumination intensities do add up to a small value in wide spectral ranges. This could be interpreted as a "neutral element" with respect to the measurement uncertainty at the present low power densities and to the fluctuations in the emission power. In contrast, there are regions of distinct deviations from a theoretically expected flat line, particularly in the case of peaks in the IR / IC range. The sum shows that the two spectra are slightly offset from each other. If this displacement, which is presumably due to the modification of the experiment between the measurements, is reduced by shifting one of the spectra by one measuring point, the sum (black-solid line) results in a significantly better compensation of the two spectra.

Remaining deviations in the peak regions can be largely attributed to slight differences in the position of the sensor at corresponding measuring points in the two spectra along the steep slopes of the IR or IC peaks. Furthermore, minor deviations are caused by the profile changes of the spectra due to the subtraction of the spectral background if it is not homogeneous in the full measuring-range.

However, the subtraction of the background is not necessary to prove the inversion of the two spectra. For then, in the inversion equation the total irradiance $E_{tot}$ is the constant to which the two spectra add up:

$$E(x) + \bar{E}(x) = E_{tot} \quad \forall x \quad (2)$$

In this form, the inversion equation can be interpreted as a conservation theorem: the two spectral irradiances sum up to the beam irradiance $E_{tot}$ of the light source. It can theoretically be argued that this is exactly the case if absorptions do not occur at any point in the experiment. In this case the optical experiment can be treated as conservative system with respect to the radiation energy (Rang 2015; Rang et al. 2017). In the present case, this condition was fulfilled by the mirror coating of the slit aperture. The experiment thus came close to being a non-dissipative system.

This consideration justifies the equal treatment of the two spectra with respect to their different spectral backgrounds: it corresponds to the subtraction of the constant total irradiance in the inversion equation (2). The irradiances can thus not only are interpreted as effective spectral components on top of their respective backgrounds, but the inversion equation passes over to the symmetrical form (1) of a mathematical inversion, which is free of additive constants.

**Conclusions**

1. The existence of the investigated inverse spectral regions *ultra-yellow* (UY) and *infra-cyan* (IC) demonstrates empirically that the inverse spectrum is not merely a visual phenomenon but should be seen as the consequence of radiation energy conservation in spectral experiments.

2. If the data of inverse spectra are evaluated in the same way, then a spectrum with only positive spectral components (ordinary spectrum) as well as an inverse spectrum with equal absolute but negative spectral components are obtained, irrespective of the measured radiometric quantities, the spectral range and the illumination used.

3. In the past, the IR range has been explored through temperature measurements. This has recently led to the question of whether "negative temperatures" can be measured in the inverted IR range (Müller 2015). Since the temperature measurement, like the measurement presented here, is based on power densities, this assumption can be confirmed. If temperature is measured in mutually inverse spectra and the measured values are converted into differences to the temperature of the respective backgrounds, a spectrum with only positive values (ordinary spectrum) and a spectrum with only negative values (inverse spectrum) always results.

4. In the field of classical spectroscopic methods, a spectrum cannot be empirically prioritized in favour of its inverse counterpart: they are spectroscopically equivalent and this equivalence is guaranteed by the energy conservation of the radiation in the spectra pair.

**Aknowledgements**

We thank Olaf Müller for encouraging us in the above investigation, Johannes Kühl and Oliver Passon for constructive discussions, David Auerbach and Laura Liska for language proofreading and the DAMUS-DONATA e.V. for the financial support that enabled us to realize the project.